\documentclass[aps,twocolumn,groupedaddress,showpacs,floatfix]{revtex4}

\usepackage{graphicx}
\begin{document}

\title{High temperature gate control of quantum well spin memory}

\author{O. Z. Karimov}
\author{G. H. John}
\author{R. T. Harley}
\affiliation{Department of Physics and Astronomy, University of
Southampton, Southampton, SO17 1BJ, UK}

\author{W. H. Lau }
\author{M. E. Flatt\'{e}}
\affiliation{Department of Physics and Astronomy, University of
Iowa, Iowa City, IA 52242, USA}

\author{M. Henini}
\affiliation{School of Physics and Astronomy, University of
Nottingham, Nottingham NG7 2RD, UK}

\author{R. Airey}
\affiliation{EPSRC Central Facility for III-V Semiconductors,
Department of Electronic and Electrical Engineering, University of
Sheffield, Sheffield S1 3JD, UK}

\date{\today}

\begin{abstract}
Time-resolved optical measurements in (110)-oriented GaAs/AlGaAs
quantum wells show a ten-fold increase of the spin-relaxation rate
as a function of applied electric field from 20 to 80 kV cm$^{-1}$
at 170 K and indicate a similar variation at 300 K, in agreement
with calculations based on the Rashba effect. Spin relaxation is
almost field-independent below 20 kV cm$^{-1}$ reflecting quantum
well interface asymmetry. The results indicate the achievability
of voltage-gateable spin-memory time longer than 3 ns
simultaneously with high electron mobility.
\end{abstract}


\maketitle The longest possible spin memory and an ability to
control the orientation or relaxation of non-equilibrium spin
populations in semiconductor quantum wells (QWs) via an applied
gate voltage will be the key to many spintronic applications.
Recent experiments on (110)-oriented III-V QWs have demonstrated a
predicted \cite{Oh99,DK86} dramatic increase of spin memory at
room temperature, up to 20 ns in an $n$-type GaAs/AlGaAs QW
\cite{Ada01}, but although various possible approaches exist, gate
control has hitherto been more elusive. The dependence of carrier spin
relaxation rates on gate-injected electron or hole concentration
in QWs has been investigated by optical techniques
\cite{Sn90,San01}. A second approach exploits the electric field
dependence of the electron $g$-factor in parabolic potential wells
to control spin precession in applied magnetic field \cite{Sal02}.
Similarly gated ferromagnetism in transition-metal-doped III-V's
\cite{Oh00} may allow control of spin orientation without an applied
magnetic field. A third proposal is to manipulate the conduction
band spin-splitting by an applied electric field through the Rashba
effect \cite{Ra60}. This permits gate control of the
exponential spin relaxation at high temperatures where electrons
experience strong scattering, the so called collision-dominated
regime, \cite{Lau02} or even coherent spin reorientation under
collision-free conditions at very low temperatures \cite{DD90}.
For the collision-dominated case this last approach has the strong
advantage of applicability at room temperature.

In this paper we report a time-resolved optical investigation of
electron spin relaxation at high-temperatures in undoped
(110)-oriented GaAs/AlGaAs QWs. We confirm the existence of a
100-fold increase of spin memory compared to (001)-oriented QWs at
300 K \cite{Oh99} but we now demonstrate a ten-fold variation of
the spin relaxation rate with the application of a modest electric
field. This is in accord with theoretical expectations
\cite{Lau02} for the Rashba effect and strongly indicates the
mechanism which currently limits the spin memory in (110)-oriented
samples in zero field.

\begin{figure}
\includegraphics[scale=0.45]{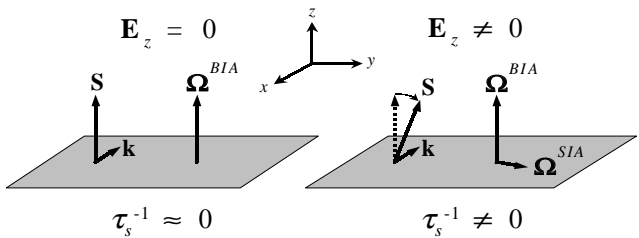}
\caption{\label{Fig1}Precession of photo-injected electron spin
population $(\bf{S})$ in a (110)-oriented QW (shown shaded) due to
conduction band spin-splitting. For zero growth-axis electric
field $({\bf{E}}_z=0)$ precession vector ${\bf{\Omega}}^{BIA}$ is
almost along the growth axis for all electron wavevectors
$(\bf{k})$ but for ${\bf{E}}_z\neq 0$ an additional in-plane
Rashba component ${\bf{\Omega}}^{SIA}$ causes precession of
$\bf{S}$ away from the growth axis. In collision-dominated
conditions this leads via Eq. (\ref{Eq1}) to a longitudinal spin
relaxation rate $\tau_s^{-1} \approx 0$ for ${\bf{E}}_z = 0$ and
to $\tau_s^{-1}\neq 0$ for finite ${\bf{E}}_z$.}
\end{figure}

The spin relaxation in our experiments can be interpreted on the
basis of the D'yakonov, Perel' and Kachorovskii (DPK)
\cite{DK86,DP71} mechanism of spin relaxation, as refined in a
non-perturbative approach \cite{Lau02,Fl01,Fl02}, which dominates
zero-field spin dynamics in GaAs/AlGaAs QW systems
\cite{Fl02,Br98,Ter99,Ma01}. In this model, spin reorientation is
driven by precession of the individual electron spin vectors
induced by the effective magnetic field represented by the
conduction band spin splitting, which results from spin-orbit
coupling and lack of inversion symmetry. The corresponding Larmor
precession vector $\bf{\Omega}(k)$ varies in magnitude and
direction according to the electron's wavevector $\bf{k}$ so that
scattering results in a randomly fluctuating precession vector
\cite{DK86,DP71,JR95}. In the collision-dominated regime,
$\left<|{\bf{\Omega}}|\right>\tau_p^*\ll 1$ with
$\left<|\bf{\Omega}|\right>$ the average precession frequency and
$\tau_p^*$ the momentum scattering time of an electron, spin
relaxation is inhibited by scattering and has rate
\cite{DK86,DP71,Fl02}
\begin{equation}\label{Eq1}
  \tau_s^{-1}=\left<\Omega^2\right>\tau_p^\ast.
\end{equation}
The first factor in this expression is determined by the vector
$\bf{\Omega}$, which, in a QW, has three contributions,
${\bf{\Omega}}^{SIA}$, ${\bf{\Omega}}^{BIA}$ and
${\bf{\Omega}}^{NIA}$ \cite{Fl02}. The natural interface asymmetry
component ${\bf{\Omega}}^{NIA}$ does not occur in GaAs/AlGaAs QWs
and is not considered further here. The most interesting for
gating spin relaxation is ${\bf{\Omega}}^{SIA}$, the Rashba or
structural inversion asymmetry (SIA) term. It may arise from some
built-in asymmetry of the structure or be induced by an externally
applied odd parity perturbation such as an electric field. For
field ${\bf{E}}_z$ along the growth axis ($z$) it has the form
(${\bf{E}}_z$$\times\bf{k}$) \cite{Ra60,Fl02} and so is oriented
in the QW plane for all in-plane $\bf{k}$. This induces precession
of the electron spin away from the $z$-axis and therefore
contributes a term to DPK spin relaxation along the axis, which
can be varied with applied electric field \cite{Lau02}. However
except in (110)-oriented QWs the effect will be small because
${\bf{\Omega}}^{BIA}$, which is due to bulk inversion asymmetry
(BIA) of the zinc blende structure and is only weakly
field-dependant, is either wholly or partially in the QW plane and
so swamps ${\bf{\Omega}}^{SIA}$ \cite{Lau02,DD90,JR95}. For the
special case of (110)-oriented QWs, illustrated in Fig.
\ref{Fig1},
${\bf{\Omega}}^{BIA}$ lies approximately along the growth axis for
all $\bf{k}$, causing only weak spin relaxation along the growth
axis and thereby leading to greatly enhanced spin memory, as
predicted by D'yakonov and Kachorovskii \cite{DK86} and confirmed
experimentally by Ohno \textit{et al.} \cite{Oh99}. Spin
relaxation along the growth axis therefore becomes more sensitive
to ${\bf{\Omega}}^{SIA}$ and should be significantly changed by
the application of an external electric field \cite{Lau02}, an
important possibility, which we investigate here.

Equation (\ref{Eq1}) also shows that the spin relaxation rate
scales with $\tau_p^*$ in a way analogous to motional narrowing.
For undoped QWs, $\tau_p^*$ is closely related to the momentum
relaxation time $\tau_p$ which determines the electron mobility
\cite{DK86,DP71,Fl02}. This means that long spin memory and high
mobility are, in general, mutually exclusive in low-doped $n$-type
QWs but may be simultaneously achievable in the special case of
(110)-oriented QWs.

\begin{figure}
\includegraphics[scale=0.45]{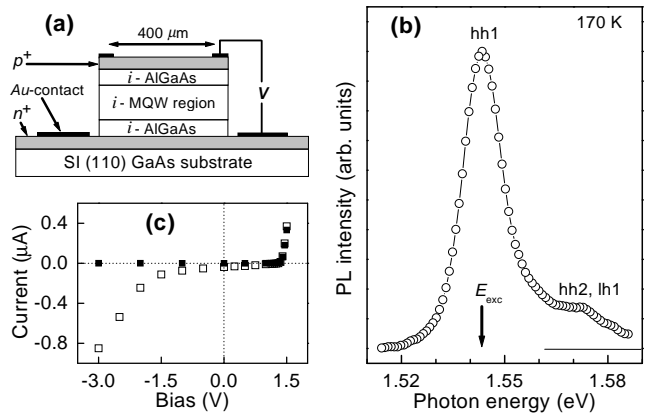}
\caption{\label{Fig2}(a) Cross-section of a 400 $\mu$m
(110)-oriented $pin$ mesa device. Pump (circular polarised) and
probe (linear polarised) beams for probing spin dynamics are
focused on the top of the mesa. (b) PL spectrum of device for
applied bias V = 0 volts at 170 K; $E_{exc}$ indicates excitation
energy for pump and probe measurements. (c) $I$-$V$
characteristics of mesa device at 170 K in dark (solid squares)
and under 350 $\mu$W laser illumination (open squares) at peak of
hh1 luminescence.}
\end{figure}

Samples from three different wafers have been studied each grown
on a semi-insulating (SI) GaAs substrate and each containing
twenty 7.5 nm QWs separated by 12 nm barriers with aluminium
fraction 0.4. Two wafers consisted only of undoped layers on
(001)- and (110)-oriented substrates respectively. The third wafer
was a $pin$ structure on a (110)-oriented substrate with the
undoped QWs grown between two 0.1 $\mu$m undoped buffer layers of
AlGaAs and with layers of 2$\times10^{18}$ cm$^{-3}$ $n^+$ doped
GaAs and of 1.2$\times10^{18}$ cm$^{-3}$ $p^+$ doped AlGaAs
respectively below and above. The orientation of the
(110)-oriented substrates was accurate to $\pm 0.5^{\circ}$.
Measurements were carried out at 300 K on all three wafers as
grown and the $pin$ wafer was also processed into 400 $\mu$m mesa
devices (Fig. \ref{Fig2}a) and used for measurements in variable
electric field at 170 K. Although the sequence of layers in the
samples does not have inversion symmetry, for flat-band conditions
the QWs and barriers possess local inversion symmetry where the
confined electron wavefunctions have significant density. The
doped layers of the $pin$ structure result in a built-in electric
field ${\bf{E}}_z$ of about 25 kV cm$^{-1}$. Figure \ref{Fig2}b
shows the photoluminescence (PL) spectrum of one of the mesa
structures for zero applied bias at 170 K. The width, at 10 K, of
the excitonic recombination line was $\sim$6.5 meV in each of our
(110)-oriented wafers and 1.8 meV in our (001)-oriented wafer.
Since we expect similar fluctuations in QW width for the two sets
of multi-quantum wells (MQWs), this suggests that the interfaces
of the (110)-oriented QWs are less perfect than for (001) growth
and could be further improved by variation of growth conditions.

Electron spin evolution was obtained from the change of intensity
$(\Delta R)$ and polarisation rotation $(\Delta \theta)$ of
optical probe pulses reflected at near normal incidence from the
sample surface along the growth axis, at a variable delay
following excitation by 10 times more intense, nearly collinear
circularly polarised pump pulses \cite{DD90,Br98,Ma01}. The pulses
were of 2 ps duration from a mode-locked Ti-sapphire laser tuned
to the peak of the first electron-heavy-hole PL transition (Fig.
\ref{Fig2}b) giving excitation density $\sim$10$^9$ cm$^{-2}$.
Figure \ref{Fig2}c shows the $I$-$V$ characteristics of the device
in the dark and illuminated with experimental laser intensity,
$\sim$350 $\mu$W. Such low powers are necessary to avoid screening
of the electric field in the MQW region. Since this incident
intensity could generate a photocurrent up to 30 $\mu$A, the
magnitude of the negative bias current shows that in fact a very
small fraction, $\sim$2$\%$, of the photo-carriers is swept out by
the bias but even this may be exaggerated because the rapid
increase of current below $-$1.5 volts indicates avalanche
multiplication. The absorbed pump photons thus generate `cold'
excitons which dissociate into free carriers on a sub-picosecond
timescale with spins polarised along the growth axis and which
remain confined in the QWs. The hole spins rapidly relax whereas
the electron spin-relaxation is much slower \cite{Br98,Ma01}. On a
timescale longer than $\sim$1 ps, $\Delta R$ gives a measure of
the population of photoexcited carriers and $\Delta \theta$ a
measure of the $z$-component of electron spin so that their ratio
gives the pure longitudinal spin dynamics of the electrons
\cite{Br98,Ma01}.
\begin{figure}
\includegraphics[scale=0.45]{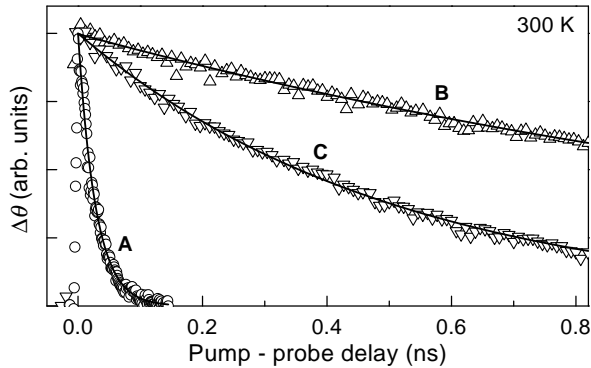}
\caption{\label{Fig3} $\Delta \theta$ signals for three as-grown
wafers each containing twenty undoped 7.5 nm GaAs/AlGaAs QWs at
300 K; A, (001)-oriented substrate, all layers undoped; B, as A
but on (110)-oriented substrate; C, as B but with $n^+(p^+)$ doped
layers below (above) the QWs giving built-in electric field
$E_z\approx25$ kV cm$^{-1}$. Corresponding values of $\tau_s$ are
$32\pm1$ ps, $3.5\pm0.2$ ns and $0.85\pm0.02$ ns respectively.}
\end{figure}

Figure \ref{Fig3} shows the observed $\Delta \theta$ signals for
the three as-grown wafers at 300 K. There is a dramatic difference
between the decay for the (001)-oriented wafer (A) and the undoped
(110)-oriented wafer (B), which essentially reproduces the
findings of Ohno \textit{et al.} \cite{Oh99}. However the decay
for the $pin$ (110)-oriented wafer (C), where there is a built-in
electric field of about 25 kV cm$^{-1}$ (at 300 K), is
significantly faster than for the undoped (110)-oriented wafer
(B). Measurements were made at various pump intensities to allow
extrapolation to zero power and when the decay of the $\Delta R$
signal is also included in each case the spin relaxation times are
found to be $32\pm1$ ps, $3.5\pm0.2$ ns and $0.85\pm0.02$ ns for
A, B and C wafers respectively. These figures are consistent with
our theoretical predictions \cite{Lau02} if we assume electron
mobilities $\sim$0.4 m$^2$ V$^{-1}$s$^{-1}$, $\sim$0.3 m$^2$
V$^{-1}$s$^{-1}$ and $\sim$0.43 m$^2$ V$^{-1}$s$^{-1}$
respectively. Although we do not have direct mobility
measurements, these are reasonable room temperature values for
such samples where optical phonon scattering and interface
roughness are likely to be dominant \cite{Br98}. Figure \ref{Fig3}
therefore gives a strong indication that there is a significant
variation of spin relaxation rate with electric field at 300 K as
predicted theoretically \cite{Lau02}.
\begin{figure}
\includegraphics[scale=0.45]{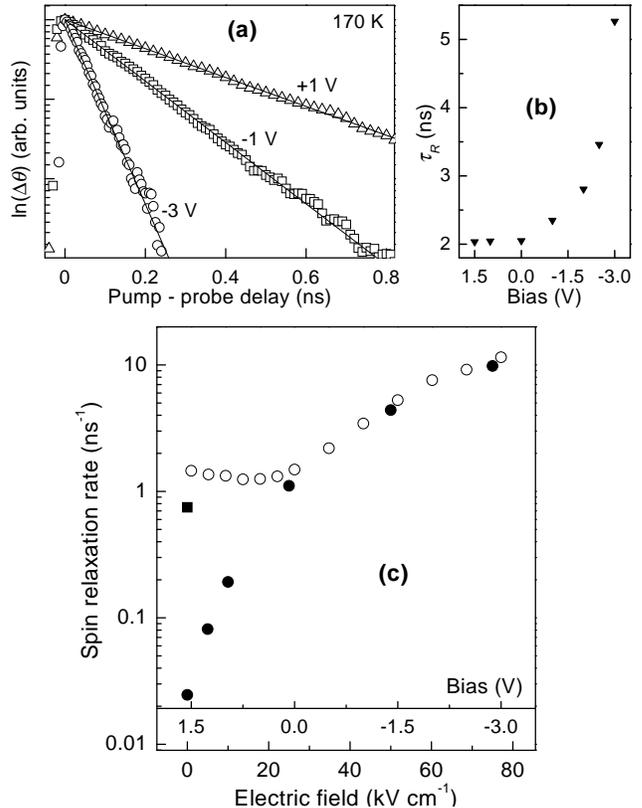}
\caption{\label{Fig4} Spin-dynamics for (110)-oriented $pin$ mesa
device at 170 K: (a) $\Delta \theta$ signals for three applied
voltages and (b) decay time for the $\Delta R$ signal showing
increase of recombination time with applied electric field. (c)
Measured spin relaxation rate vs bias voltage and corresponding
electric field (open circles) compared with calculation for a
symmetrical QW (solid circles) and for a QW with one two-monolayer
graded interface (solid square) assuming electron mobility 0.6
m$^2$ V$^{-1}$s$^{-1}$.}
\end{figure}

Measurements on the mesa devices at different bias voltages at 170
K (Fig. \ref{Fig4}a) support this conclusion (measurements at 300
K for the same range of voltages were prevented by excessive
current in reverse bias leading to destruction of the device). The
$\Delta \theta$ signals show single-exponential decay and there is
a strong variation of decay rate with voltage. The $\Delta R$
signals also decay exponentially and Fig. \ref{Fig4}b shows the
extracted decay times ($\tau_{R}$) as a function of bias. For
large negative bias $\tau_{R}$ increases significantly reflecting
the reduced electron-hole overlap at high electric fields and
showing that recombination is predominantly radiative
\cite{Pol85}.

Figure \ref{Fig4}c shows the spin relaxation rate vs bias voltage
and vs the corresponding electric field at the QWs obtained using
the layer thicknesses in the $pin$ structure and the band gap of
GaAs at 170 K. Up to 20 kV cm$^{-1}$ the spin relaxation rate is
almost constant but then increases by about a factor of 10 by 80
kV cm$^{-1}$. The solid dots are results of our non-perturbative
calculations \cite{Lau02,Fl01,Fl02} based on the DPK mechanism for
a symmetrical (110)-oriented QW. The calculations assumed electron
mobility of 0.6 m$^2$ V$^{-1}$s$^{-1}$, which is consistent with
the assumed 300 K value in the $pin$ wafer, $\sim$0.43 m$^2$
V$^{-1}$s$^{-1}$, for mobility limited by optical phonon
scattering in this temperature range. The variation for electric
fields greater than 20 kV cm$^{-1}$ (Fig. \ref{Fig4}c) is very
well fitted by the theory but at lower fields there is clearly a
contribution to the spin relaxation not included in the
calculations which predict a rate almost 100 times lower than we
have observed at zero field.

The measured decay rates in this region were found to increase
linearly with pump power, consistent with a small influence of the
Bir, Aronov and Pikus (BAP) relaxation mechanism \cite{BAP76} due
to exchange interaction with the photo-excited holes as also found
by Adachi \textit{et al.} \cite{Ada01}. This does not, however,
explain the observed discrepancy with theory since extrapolation
to zero power would only make a reduction of about 10$\%$. Nor can
it be explained in terms of the BAP mechanism with electrically
injected holes; even for flat band conditions in forward bias,
where the injected hole concentration would be greatest, the
calculated concentration is at least two orders of magnitude too
low to explain the additional spin relaxation \cite{BAP76}.
Another possible cause of the additional spin relaxation is random
built-in asymmetry in the QWs, which would give a
field-independent SIA contribution. Such asymmetry could result
from alloy fluctuations or from differences of top and bottom
interface morphology for the GaAs/AlGaAs QWs and for a given
$\bf{k}$ would generate an ${\bf{\Omega}}^{SIA}$ component varying
randomly in magnitude and orientation from point to point on a QW
with associated non-zero mean square entering Eq. (\ref{Eq1}). To
gauge this possibility, we have repeated our calculation in zero
field for a QW with one perfect interface and the other containing
a two-monolayer compositional gradient. The relaxation rate is
$\sim$30 times greater than for a perfectly symmetrical QW (solid
square in Fig. \ref{Fig4}c). Even this small perturbation of the
symmetry produces a change approaching our discrepancy between
theory and experiment supporting the idea that the spin memory is
actually limited by the sample perfection which, as indicated by
the width of luminescence lines may be somewhat inferior to that
in (001)-oriented samples. At sufficiently high reverse bias the
increase of the relaxation rate due to the applied field becomes
dominant over this random built-in asymmetry.

In conclusion, we have demonstrated that the enhanced
high-temperature electron spin memory in (110)-oriented QWs may be
varied by at least a factor 10 by application of modest gate bias
voltages. The variation is consistent with the Rashba effect and
the DPK spin-relaxation mechanism. On the basis of this mechanism,
the measurements for low electric fields indicate that a further
significant enhancement of spin memory may be achieved by
modification of the growth techniques to optimise the interface
morphology. This may be expected to give at 300 K spin memory
longer than 10 ns \cite{Lau02,Ada01} and straightforward voltage
bias control simultaneously with high electron mobility. This
combination of properties will facilitate a variety of spintronic
devices.

\begin{acknowledgments}
We wish to thank Professor E. L. Ivchenko, Mr M. M. Glazov and Dr.
K. V. Kavokin for many illuminating discussions. This work was
supported by EPSRC and DARPA/ARO.
\end{acknowledgments}


\begin{thebibliography}{99}

\bibitem[1]{Oh99}
Y. Ohno \textit{et al.}, Phys. Rev. Lett. \textbf{83}, 4196
(1999).

\bibitem[2]{DK86}
M. I. D'yakonov and V. Yu. Kachorovskii, Sov. Phys. Semicond.
\textbf{20}, 110 (1986).

\bibitem[3]{Ada01}
T. Adachi \textit{et al.}, Physica E \textbf{10}, 36 (2001).

\bibitem[4]{Sn90}
M. J. Snelling \textit{et al.}, J. Luminescence \textbf{45}, 208
(1990).

\bibitem[5]{San01}
J. S. Sandhu \textit{et al.}, Phys. Rev. Lett. \textbf{86}, 2150
(2001).

\bibitem[6]{Sal02}
G. Salis \textit{et al.}, Nature (London) \textbf{414}, 619
(2002); for review see: D. D. Awschalom, D. Loss and N. Samarth,
eds., \textit{Semiconductor Spintronics and Quantum Computation}
(Springer, 2002), chap. 5: \textit{Optical Manipulation of Spin
Coherence in Semiconductors} by D. D. Awschalom and N. Samarth.

\bibitem[7]{Oh00}
H. Ohno \textit{et al.}, Nature (London) \textbf{408}, 944 (2000).

\bibitem[8]{Ra60}
E. I. Rashba, Sov. Phys. Solid State \textbf{2}, 1109 (1960); Y.
A. Bychkov and E. I. Rashba, J. Phys. C \textbf{17}, 6039 (1984).

\bibitem[9]{Lau02}
W. H. Lau and M. E. Flatt\'{e}, J. Appl. Phys. \textbf{91}, 8682
(2002).

\bibitem[10]{DD90}
S. Datta and B. Das, Appl. Phys. Lett. \textbf{56}, 665 (1990); M.
A. Brand \textit{et al.}, Phys. Rev. Lett., \textbf{89}, 236601
(2002).

\bibitem[11]{DP71}
M. I. D'yakonov and V. I. Perel', Sov. Phys. JETP \textbf{33},
1053 (1971).

\bibitem[12]{Fl01}
W. H. Lau, J. T. Olesberg and M. E. Flatt\'{e}, Phys. Rev. B
\textbf{64}, 161301 (2001).

\bibitem[13]{Fl02}
for review see: D. D. Awschalom, D. Loss and N. Samarth, eds.,
\textit{Semiconductor Spintronics and Quantum Computation}
(Springer, 2002), chap. 4: \textit{Spin Dynamics in
Semiconductors} by M. E. Flatt\'{e}, J. M. Byers and W. H. Lau.

\bibitem[14]{Br98}
R. S. Britton \textit{et al.}, Appl. Phys. Lett. \textbf{73}, 2140
(1998).

\bibitem[15]{Ter99}
R. Terauchi \textit{et al.}, Jpn. J. Appl. Phys. \textbf{38}, 2549
(1999).

\bibitem[16]{Ma01}
A. Malinowski \textit{et al.}, Phys. Rev. B \textbf{62}, 13034
(2001); R. T. Harley, O. Z. Karimov and M. Henini, J. Phys. D, in
press.

\bibitem[17]{JR95}
B. Jusserand \textit{et al.}, Phys. Rev. B \textbf{51}, 4707
(1995).

\bibitem[18]{Pol85}
H. -J. Polland \textit{et al.}, Phys. Rev. Lett. \textbf{55}, 2610
(1985).

\bibitem[19]{BAP76}
G. L. Bir, A. G. Aronov and G. E. Pikus, Sov. Phys. JETP
\textbf{42}, 705 (1976); G. Fishman and G. Lampel, Phys. Rev. B
\textbf{16}, 820 (1977); for review see: F. Meier and B. P.
Zakharchenya, eds., Optical Orientation (North-Holland, 1984).

\end{thebibliography}
\end{document}